\documentclass[final]{article}
\usepackage{physics}
\usepackage[final]{graphicx}
\usepackage{graphicx}
\usepackage{amssymb}
\usepackage{rotating}
\usepackage{pdflscape}
\usepackage[english]{babel}
\usepackage{color}
\usepackage{bigintcalc}
\def\beq{\begin{eqnarray}}
\def\eeq{\end{eqnarray}}
\usepackage{amsfonts}

\usepackage{seqsplit}

\newtheorem{rule-of-thumb}[theorem]{Definition} 

\usepackage{amsmath}

\makeatletter
\long\def\dddddot#1{%
  {\mathop {#1}\limits ^{\vbox to-1.4\ex@ {\kern -\tw@ \ex@ \hbox {\normalfont .....}\vss }}}%
}
\long\def\multidots#1#2{%
  \count@=0
  {{\mathop {#2}\limits ^{\vbox to-1.4\ex@ {\kern -\tw@ \ex@ \hbox {\normalfont %
  \loop%
  \ifnum#1>\count@%
  .%
  \advance\count@ by1%
  \repeat%
  }\vss }}}}%
}
\makeatother

\begin{document}


\title{Quantum particles in a suddenly moving localized potential}

\author{
Miguel Ahumada-Centeno \\
Facultad de Ciencias,  Universidad de Colima,\\
Bernal D\'{i}az del Castillo 340, Colima, Colima, Mexico  \\
mahumada0@ucol.mx 
\and
Paolo Amore \\
Facultad de Ciencias, CUICBAS, Universidad de Colima,\\
Bernal D\'{i}az del Castillo 340, Colima, Colima, Mexico  \\
paolo.amore@gmail.com 
\and
Francisco M Fern\'andez \\
INIFTA,  Division Quimica Teorica, \\
Blvd. 113 S/N, Sucursal 4, Casilla de Correo 16, 1900 La Plata, Argentina \\
fernande@quimica.unlp.edu.ar 
\and
Jesus Manzanares \\
Universidad de Sonora, Mexico \\
new.jmanza@gmail.com 
}

\maketitle

\begin{abstract}
  We study the behavior of a quantum particle, trapped in localized potential, when the trapping potential starts suddenly to move with constant velocity. In one dimension we have reproduced the results obtained 
by Granot and Marchewka, Ref.~\cite{Granot09},  for an attractive delta function, using an approach based on  a spectral decomposition, rather than on the propagator. 
We have also considered the cases of P\"oschl-Teller and simple harmonic oscillator potentials (in one dimension) and to the hydrogen atom (in three dimensions). In this last case we have calculated explicitly the leading contribution to the ionization probability for the hydrogen atom due to a sudden movement.
\end{abstract}

\section{Introduction}
\label{Intro}

In a recent paper, Ref.~\cite{Granot09}, Granot and Marchewka have studied the interesting problem of determining the behavior
of  a quantum particle trapped in a localized potential, when the potential suddenly starts to move at constant speed at $t=0$. These
authors used an attractive Dirac delta potential in one dimension to model the problem, calculating exactly the probability that
the particle remains confined to the moving potential or that it remains in the initial position. In addition to these two possibilities, they also 
observed  the probability that the particle moves at twice the speed of the potential: for an observer sitting in the rest frame
of the potential at $t>0^+$ this phenomenon can be interpreted as a quantum reflection of a particle moving to the left with speed $-v$
from the well. 

The technological advances are making increasingly realistic the scenarios where a quantum particle, such as an atom, can be 
trapped by suitable attractive potential (corresponding for example to a tip of a needle in a scanning tunneling microscope or a highly focused laser beam
in an optical tweezer) and thus be relocated in a different region~\cite{Ashkin86,Block90,Phillips14,Moffitt08}. As remarked by Granot and Marchewka, the quantum nature of this process leads to surprising results, as the possibility that the particle moves at twice the speed of the needle. 

The present paper has two different goals: first, to reproduce the analysis of Ref.~\cite{Granot09} using a more direct approach based 
on a spectral decomposition rather than on the propagator; second, to extend the analysis to a  wider class of problems, in one and 
three dimensions.

The paper is organized as follows: in Section \ref{spectral} we discuss the general framework, using spectral decomposition; in 
Section \ref{appl} we consider several examples of potentials, with spectra which can be either mixed or discrete, and calculate
{\sl explicitly} the relevant probabilities for each case; finally in Section \ref{concl} we draw our conclusions.

\section{Spectral decomposition}
\label{spectral}

Our starting point is the time dependent Schr\"odinger equation (TDSE)
\begin{eqnarray}
i \hbar \frac{\partial \Psi}{\partial t} = -\frac{\hbar^2}{2m} \Delta \Psi + V(\vec{r}-\vec{v} t) \Psi(\vec{r},t)
\label{eq_tdse}
\end{eqnarray}
where $V(\vec{r}-\vec{v} t)$ is a potential moving with velocity $\vec{v}$.

Let us define $\vec{\xi} (\vec{r},t) \equiv \vec{r}-\vec{v} t$ and let $\phi(\vec{\xi})$ be the solution of the time independent
Schr\"odinger equation (TISE)
\begin{eqnarray}
- \frac{\hbar^2}{2m} \Delta_{\xi} \phi  + V(\vec{\xi}) \phi(\vec{\xi}) = E \phi(\vec{\xi})
\label{eq_tise}
\end{eqnarray}

It can be easily verified that 
\begin{eqnarray}
\Psi(\vec{r},t) = e^{\frac{i m \vec{v} \cdot \vec{r}}{\hbar} -\frac{i m \vec{v}^2 t}{2\hbar}} \phi(\vec{\xi}) e^{-\frac{i E t}{\hbar}}
\label{subst}
\end{eqnarray}
is a solution to the time-dependent Schr\"odinger equation (\ref{eq_tdse}).

The physical situation studied by Granot and Marchewka amounts to having a quantum particle at the initial time in the ground state
of the static potential and determining the probability at later times $t>0$ that the particle can be found in any of the modes
of the moving potential. For simplicity we will denote as $\psi_n(\vec{r},t)$ and $\psi_{\vec{k}}(\vec{r},t)$ the eigenmodes of the
moving potential and $\phi_n(\vec{r})$ and  $\phi_{\vec{k}}(\vec{r})$ the eigenmodes of the static potential, where $n$ and $\vec{k}$ refer to bound an continuum states, respectively. 
We also call $\Phi(\vec{r})$ the initial wave function at $t=0$ (for the specific case studied in Ref.~\cite{Granot09} this is 
the wave function of the ground state).

Let $\Psi(\vec{r},t)$ be the wave function solution to the TDSE with a moving potential at $t>0$, subject to the condition
$\Psi(\vec{r},0) = \Phi(\vec{r})$; this wave function can be naturally decomposed in the basis of the moving potential as
\begin{eqnarray}
\Psi(\vec{r},t) = \sum_{n} a_n \psi_n(\vec{r},t) + \int \frac{d^3k}{(2\pi)^3}  b(\vec{k}) \psi_{\vec{k}}(\vec{r},t)
\label{eq_arb}
\end{eqnarray}
where, using eq.~(\ref{subst}),
\begin{eqnarray}
a_n &=& \int d^3r \psi_n^\star(\vec{r},0) \Phi(\vec{r}) = \int d^3r e^{- i \frac{m \vec{v} \cdot \vec{r}}{\hbar}}
\phi_n^\star(\vec{r},0) \Phi(\vec{r})  \nonumber \\
b({\vec{k}}) &=& \int d^3r \psi_{\vec{k}}^\star(\vec{r},0) \Phi(\vec{r}) = \int d^3r e^{- i \frac{m \vec{v} \cdot \vec{r}}{\hbar}}  \phi_{\vec{k}}^\star(\vec{r},0) \Phi(\vec{r})  \nonumber \ .
\end{eqnarray}

In this way $\sum_n |a_n|^2$ is the probability that the particle remains in a bound state when the potential starts moving, whereas
$\int \frac{d^3k}{(2\pi)^3} |b(\vec{k})|^2$ is the probability that the particle will escape to the continuum~\footnote{We are discussing here the three dimensional case, but the modifications for the one and two dimensional cases are straightforward.}.

Calling $N$ the number of bound states of the potential, we define the physical amplitudes 
\begin{eqnarray}
\mathcal{Q}_{ij}(\vec{v}) &\equiv& \int_{-\infty}^{\infty} e^{- i m \vec{v} \cdot \vec{r} /\hbar} \left[\phi_j(\vec{r})\right]^\star \phi_i(\vec{r}) d^3 r \nonumber \\
\mathcal{P}_{i}(\vec{k},\vec{v}) &\equiv& \int_{-\infty}^{\infty} e^{- i m \vec{v} \cdot \vec{r} /\hbar} \left[\phi(\vec{k},\vec{r})\right]^\star \phi_i(\vec{r}) d^3r \nonumber \\
\tilde{\mathcal{P}}_{i}(\vec{k},\vec{v}) &\equiv& \int_{-\infty}^{\infty} e^{- i m \vec{v} \cdot \vec{r} /\hbar}  
\phi_i^\star(\vec{r}) \phi(\vec{k},\vec{r}) d^3r = \tilde{\mathcal{P}}_{i}^\star(\vec{k},-\vec{v})
\nonumber 
\end{eqnarray}

The wave function at later times for a particle initially in the ith state of the static potential will then be
\begin{eqnarray}
\Psi(\vec{r},t) = \sum_{j=1}^N \mathcal{Q}_{ij} \psi_j(\vec{r},t) +  \int_{-\infty}^\infty \mathcal{P}_{i}(\vec{k},v) \psi(\vec{k},\vec{r},t) \frac{d^3k}{(2\pi)^3}
\label{eq_psit}
\end{eqnarray}

Clearly $|\mathcal{Q}_{ij}|^2$ represents the probability that the particle, initially in the $ith$ bound state ends up in the
$jth$ bound state; similarly $|\mathcal{P}_i(\vec{k},v)|^2 d^3 k/(2\pi)^3$ represents the probability that the particle initially in the ith bound state ends up in the continuum with momentum in an infinitesimal volume about $\hbar \vec{k}$.

The conservation of total probability requires
\begin{eqnarray}
\sum_{j=1}^N |\mathcal{Q}_{ij}|^2 + \int_{-\infty}^{\infty} |\mathcal{P}_i(\vec{k},v)|^2 \frac{d^3k}{(2\pi)^3} = 1
\end{eqnarray}

Similarly we define the amplitude
\begin{eqnarray}
\mathcal{R}(\vec{k},\vec{k}',\vec{v}) &\equiv& \int_{-\infty}^{\infty} e^{- i m \vec{v}\cdot \vec{r} /\hbar} \left[\phi(\vec{k}',\vec{r})\right]^\star \phi(\vec{k},\vec{r}) d^3r \nonumber 
\end{eqnarray}
which is related to the probability that a particle initially with momentum $\hbar \vec{k}$ ends up in a state with momentum $\hbar \vec{k}'$.

Of course these arguments can be generalized to the case that the initial wave function is not a stationary state of
the static problem. In this case the time dependent wave function is 
\begin{eqnarray}
\Psi(\vec{r},t) = \sum_{j=1}^N \bar{\mathcal{Q}}_{j} \psi_j(\vec{r},t) +  \int_{-\infty}^\infty \bar{\mathcal{P}}(\vec{k},v) \psi(\vec{k},\vec{r},t) \frac{d^3k}{(2\pi)^3}
\label{eq_psit2}
\end{eqnarray}
where
\begin{eqnarray}
\bar{\mathcal{Q}}_j &\equiv& \int e^{- i m \vec{v} \cdot \vec{r}/\hbar} \left[\phi_j(\vec{r})\right]^\star \Phi(\vec{r}) d^3r \nonumber \\
&=&  \sum_{i=1}^N a_i \mathcal{Q}_{ij} +  \int b(\vec{k}) \tilde{\mathcal{P}}_j(\vec{k},\vec{v}) \frac{d^3k}{(2\pi)^3}  \nonumber \\
\bar{\mathcal{P}}(k',v) &\equiv& \int e^{- i m \vec{v} \cdot \vec{r} /\hbar} \left[\phi(\vec{k}',\vec{r})\right]^\star \Phi(\vec{r}) d^3r  \nonumber 
\end{eqnarray}
of which Eq.(\ref{eq_psit}) is a special case.

\section{Applications}
\label{appl}

In this section we apply our general discussion to different examples.

\subsection{Attractive Dirac delta potential} 
\label{delta}

We consider the attractive Dirac delta potential studied in Ref.~\cite{Granot09}:
\begin{eqnarray}
V(x,t)=\left\{ \begin{array}{ccc}
-\gamma \delta(x) & , & t \leq 0 \\
-\gamma \delta(x-vt) & , & t > 0 \\
\end{array}\right.
\end{eqnarray}

To implement the procedure explained in Section \ref{spectral} we first write explicitly 
the eigenfunctions of the static potential, $V(x,0)= -\gamma \delta(x)$, reported in Ref.~\cite{Blinder88}
\begin{eqnarray}
\phi_0(x) &=& \sqrt{\beta } e^{-\beta  \left| x\right| } 
\label{eq_bs} \\
\phi_p^{(e)}(x) &=& \frac{\sqrt{2} (p \cos (p x)-\beta  \sin (p \left| x\right| ))}{\sqrt{\beta ^2+p^2}} 
\label{eq_even} \\ 
\phi_p^{(o)}(x) &=& \sqrt{2} \sin (p x) 
\label{eq_odd} 
\end{eqnarray}
where $\beta = m \gamma/\hbar^2$. Damert \cite{Damert75} and Patil \cite{Patil00} have proved the completeness 
of the set of the energy eigenfunctions.
Note that the spectrum of $V(x,0)$ is mixed with a single bound state of energy $E_0 = - \frac{\gamma ^2 m}{2 \hbar ^2}$.

The orthonormality relations for this set of functions are
\begin{eqnarray}
\int_{-\infty}^{\infty} \phi_0^\star(x) \phi_0(x) dx &=& 1 \nonumber \\
\int_{-\infty}^{\infty} \phi_0^\star(x)  \phi_p^{(e)}(x) dx &=& \int_{-\infty}^{\infty} \phi_0^\star(x)  \phi_p^{(o)}(x) dx = 0  \nonumber \\
\int_{-\infty}^{\infty} {\phi_{p'}^{(e)}}^\star(x)  \phi_p^{(e)}(x) dx &=& \int_{-\infty}^{\infty} {\phi_{p'}^{(o)}}^\star(x)  \phi_p^{(o)}(x) dx =  2 \pi \delta (p-p')  \nonumber \\
\int_{-\infty}^{\infty} {\phi_{p'}^{(o)}}^\star(x)  \phi_p^{(e)}(x) dx &=& 0  \nonumber 
\end{eqnarray}

Using the prescription (\ref{subst}) one can obtain the solutions to the time dependent problem for $t>0$ corresponding
to the static solutions of eqs. (\ref{eq_bs}), (\ref{eq_even}) and (\ref{eq_odd}):
\begin{eqnarray}
\psi_0(x,t,v) &=& \sqrt{\beta } e^{-\beta  \left| x-t v\right| +\frac{i m \left(\gamma ^2 t+v \hbar ^2 (2 x-t v)\right)}{2 \hbar^3}} 
\nonumber \\
\psi_k^{(e)}(x,t,v) &=& \frac{\sqrt{2}}{\sqrt{\beta ^2+k^2}}  e^{-\frac{i \left(k^2 t \hbar ^2+m^2 v (t v-2 x)\right)}{2 m \hbar }} 
(k \cos (k (x-t v))-\beta \sin (k \left| x-t v\right|))  \nonumber \\
\psi_k^{(o)}(x,t,v) &=& \sqrt{2} \sin (k (x-t v)) e^{-\frac{i \left(k^2 t \hbar ^2+m^2 v (t v-2 x)\right)}{2 m \hbar }}    \nonumber 
\end{eqnarray}

The initial wave function is the bound state of the static delta potential
\begin{eqnarray}
\Psi_0(x) &=& \sqrt{\beta } e^{-\beta  \left| x\right| }  \nonumber 
\end{eqnarray}
and it can be decomposed in the basis of the time-dependent potential as
\begin{eqnarray}
\Psi_0(x) = \mathcal{Q}_{11}(v) \psi_0(x,0) + \int_0^\infty \frac{dk}{2\pi}  \left[\mathcal{P}_1^{(e)}(k,v) \psi_k^{(e)}(x,0)+
\mathcal{P}_1^{(o)}(k,v) \psi_k^{(o)}(x,0)\right]  \nonumber 
\end{eqnarray}
as explained in Section \ref{spectral}.

The amplitude for transition to the bound state of the moving well is given by
\begin{eqnarray}
\mathcal{Q}_{11}(v) &=& \int_{-\infty}^\infty e^{- i m v x/\hbar} \phi_0^\star(x) \phi_0(x) dx  = \frac{4}{\theta ^2+4}
\end{eqnarray}
where $\theta \equiv \hbar v/\gamma$ is the adiabatic Massey parameter \cite{Elberfeld88,Granot09}. 

Clearly, the probability that the particle remains in the bound state of the moving well is simply given by
\begin{eqnarray}
P_{bound}= |\mathcal{Q}_{11}|^2 = \frac{16}{(\theta^2+4)^2}
\end{eqnarray}
in agreement with the eq.(13) of Ref.~\cite{Granot09}.

Similarly we can calculate the coefficients $\mathcal{P}_1^{(e)}(k,v)$ and $\mathcal{P}_1^{(o)}(k,v)$, representing the amplitudes for transitions to a state in the continuum (even and odd states, respectively)
with momentum $\hbar k$ 
\begin{eqnarray}
\mathcal{P}_1^{(e)}(k,v) 
&=& \int_{-\infty}^\infty  e^{- i m v x/\hbar}   \left[{\phi_k^{(e)}}(x)\right]^\star \phi_0(x) dx \nonumber \\
&=& \frac{4 \sqrt{2} \beta^{7/2} \theta ^2 k}{\sqrt{\beta^2+k^2} \left(\beta^4 \left(\theta ^2+1\right)^2+k^4-2 \beta ^2 \left(\theta ^2-1\right) k^2\right)} \\
\mathcal{P}_1^{(o)}(k,v)
&=& \int_{-\infty}^\infty  e^{- i m v x/\hbar}  \left[{\phi_k^{(o)}}(x)\right]^\star \phi_0(x) dx \nonumber \\
&=& -\frac{4 i \sqrt{2} \beta ^{5/2} \theta k}{\beta ^4 \left(\theta^2+1\right)^2+k^4-2 \beta ^2 \left(\theta^2-1\right) k^2}
\end{eqnarray}
where $\beta  = m v/\hbar\theta$.

As a result the probability that the particle ends up in the continuum is
\begin{eqnarray}
P_{continuum} &=& \int_{0}^\infty \left[ |\mathcal{P}_1^{(e)}(k,v)|^2 + |\mathcal{P}_1^{(o)}(k,v)|^2\right] \frac{dk}{2\pi} \nonumber \\
&=& \int_{-\infty}^\infty \left[ |\mathcal{P}_1^{(e)}(k,v)|^2 + |\mathcal{P}_1^{(o)}(k,v)|^2\right] \frac{dk}{4\pi}  \nonumber \\
&=&  \int_{-\infty}^\infty \frac{16 \beta ^5 \theta ^2 k^2 \left(\beta ^2
	\left(\theta^2+1\right)+k^2\right)}{\left(\beta^2+k^2\right) \left(\beta ^4 \left(\theta^2+1\right)^2+k^4-2 \beta ^2 \left(\theta^2-1\right) k^2\right)^2}  \frac{dk}{2\pi} \nonumber
\end{eqnarray}

A straightforward integration of this expression using the residue theorem provides the final result
\begin{eqnarray}
P_{continuum} &=& 1-\frac{16}{\left(\theta ^2+4\right)^2}
\end{eqnarray}

The total probability correctly sums to $1$:
\begin{eqnarray}
P_{total} = P_{bound}+P_{continuum} = 1 \nonumber 
\end{eqnarray}

The solution at $t >0$ is then obtained using eq.~(\ref{eq_psit}):
using this expression we were able to reproduce Fig.~3 of Ref.~\cite{Granot09} performing a numerical integration 
of this equation (notice however a typo in the second equation of (15) of Ref.~\cite{Granot09}).

\subsection{P\"oschl-Teller potentials}

The second example that we want to consider is the P\"oschl-Teller (PT) potential
\begin{eqnarray}
V(x) = -\frac{\hbar^2 \lambda  (\lambda +1)}{2 a^2 m} \ \sech^2\left(\frac{x}{a}\right)
\label{eq:poschl}
\end{eqnarray}
for which exact solutions are available.

The potential (\ref{eq:poschl}) provides a nice generalization of our discussion for the attractive delta potential, both because
it has a mixed spectrum, with $\lambda$ bound states ($\lambda$ integer), and because it is known to be reflectionless~\cite{Kay56}.

The eigenfunctions of the  bound states read
\begin{eqnarray}
\phi_j(x) = \frac{\mathcal{N}_j}{\sqrt{a}} \ P_{\lambda }^j\left(\tanh \left(\frac{x}{a}\right)\right) \ \ , \ \ j=1, \dots, \lambda
\end{eqnarray}
where $P_{\lambda}^j(x)$ is the associated Legendre polynomial; the corresponding eigenenergies are
$E_j= -\frac{j^2 \hbar ^2}{2 a^2 m}$. Here $\mathcal{N}_j$ is a (dimensionless) normalization constant.

As an example we consider the case $\lambda=1$, for which
\begin{eqnarray}
\phi_1(x) &=& -\frac{\sqrt{1-\tanh^2\left(\frac{x}{a}\right)}}{\sqrt{2} \sqrt{a}} \\
\phi_k(x) &=& \frac{\left(-\tanh \left(\frac{x}{a}\right)+i a k\right)}{1+i a k} \ e^{i k x} 
\end{eqnarray}
and assume that the particle is in the ground state of the static potential at $t=0$.

The amplitudes for the transition to the bound state and to the continuum states of the moving potential can then be
calculated explicitly as
\begin{eqnarray}
\mathcal{Q}_{11}   &\equiv& \int_{-\infty}^{\infty} e^{- i m v x /\hbar} \left[\phi_1(x)\right]^\star \phi_1(x) dx 
= \frac{1}{2} \pi  \kappa {\csch}\left(\frac{\pi  \kappa}{2}\right) \nonumber \\
\mathcal{P}_{1}(k,v) &\equiv& \int_{-\infty}^{\infty} e^{- i m v x /\hbar} \left[\phi(k,x)\right]^\star \phi_1(x) dx 
= \frac{\pi  \sqrt{a} \kappa }{\sqrt{2} (a k+i)} \ {\sech}\left(\frac{1}{2} \pi  (a k+\kappa )\right) \nonumber
\end{eqnarray}
where $\kappa \equiv a mv/\hbar$.

\begin{figure}
	\begin{center}
		\bigskip\bigskip\bigskip
		\includegraphics[width=5cm]{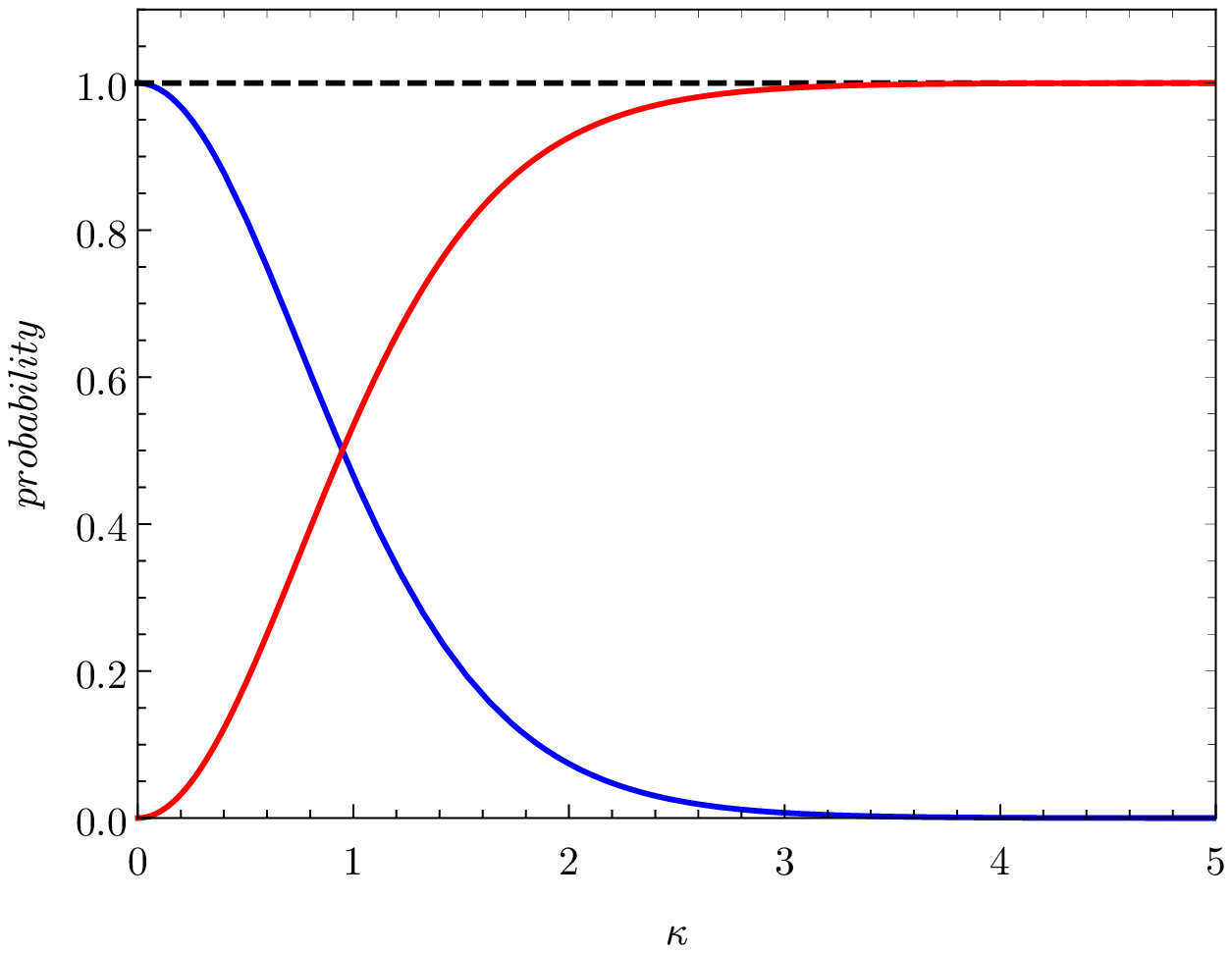} \ \ 
		\includegraphics[width=5cm]{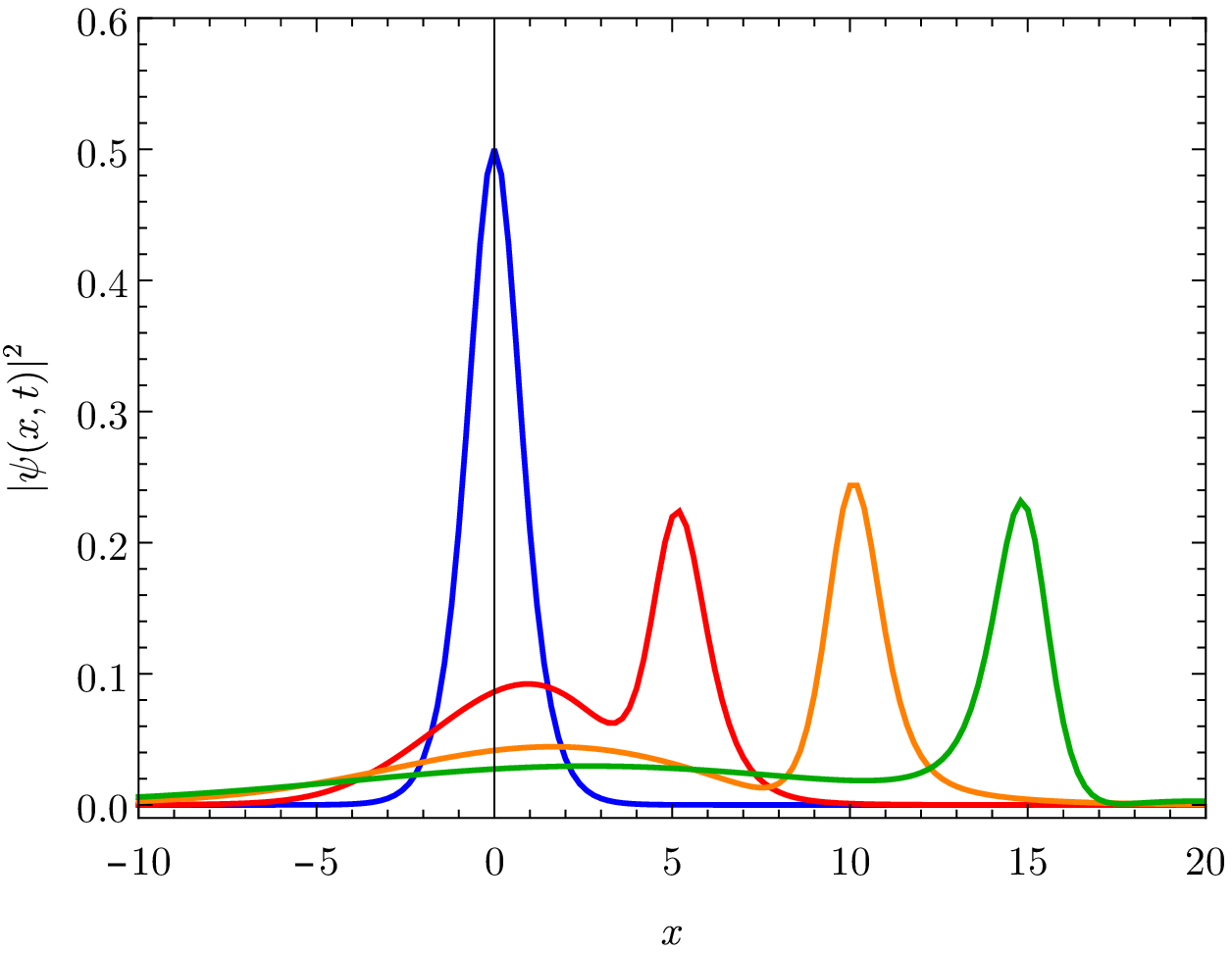} \\
		\caption{Left plot: Probability that a particle initially in the bound state of the Poschl-Teller potential with $\lambda=1$ stays
			trapped as the well starts to move with constant velocity (blue curve); the red curve is the probability that the particle ends
			up in a state of the continuum. Right plot: Probability density at four different times $t=0,5,10,15$ for a particle initially in the bound state of the static potential with $\lambda=1$.  We have used $a=m=\hbar=\kappa=1$. }
		\label{Fig_1}
	\end{center}
\end{figure}

In the left plot of Figure \ref{Fig_1} we plot the probability that a particle initially in the bound state of the Poschl-Teller potential with  $\lambda=1$ stays trapped as the well starts to move with constant velocity (blue curve),
$P_{bound}(\kappa) = | \mathcal{Q}_{11} |^2$, and the probability that the particle ends up in the continuum (red curve),
$P_{continuum}(\kappa) = \int_{-\infty}^{\infty} | \mathcal{P}_{1}(k,v) |^2 \frac{dk}{2\pi}$. The integral over momentum
is performed numerically and it is verified within the numerical accuracy that $P_{bound}(\kappa)+ P_{continuum}(\kappa) =1$.
The situation is qualitatively similar to the case treated in Ref.~\cite{Granot09}, but with $P_{bound}(\kappa)$ decaying now exponentially 
for $\kappa \gg 1$.

The wave function at $t>0$ can be then obtained as
\begin{equation}
\begin{split}
\Psi(x,t) &= \frac{1}{2} \pi  \kappa {\csch}\left(\frac{\pi  \kappa}{2}\right) \psi_1(x,t) \\ 
&+  \int_{-\infty}^\infty  \frac{\pi  \sqrt{a} \kappa }{\sqrt{2} (a k+i)} \ {\sech}\left(\frac{1}{2} \pi  (a k+\kappa )\right)  \psi(k,x,t) \frac{dk}{2\pi} \nonumber 
\end{split}
\end{equation}
where
\begin{eqnarray}
\psi_1(x,t) &=& e^{\frac{i m v x}{\hbar} -\frac{i m v^2 t}{2\hbar}} \phi_1(x-vt) e^{-\frac{i E_1 t}{\hbar}} \nonumber \\
\psi(k,x,t) &=& e^{\frac{i m v x}{\hbar} -\frac{i m v^2 t}{2\hbar}} \phi(k,x-vt) e^{-\frac{i \hbar k^2 t}{2m}}  \nonumber 
\end{eqnarray}

Since the P\"oschl-Teller potential is reflectionless we expect that the wave function would be qualitatively different from the wave function of the moving delta well, due to the absence of a peak moving with velocity $2v$.
The time evolution of a wave function for a particle initially in the bound state of the PT potential, which suddenly starts to move
with constant velocity, is displayed in the right plot Fig.~\ref{Fig_1}. We have used $a=m=\hbar=\kappa=1$; the probability density is plotted at four different times $t=0,5,10,15$. In this case it is evident the absence of the reflected wave, as expected, given the
nature of the potential. We also appreciate that the peak is moving at velocity $v$, and the corresponding 
wave function is not dispersing.

\begin{figure}
	\begin{center}
		\bigskip\bigskip\bigskip
		\includegraphics[width=5cm]{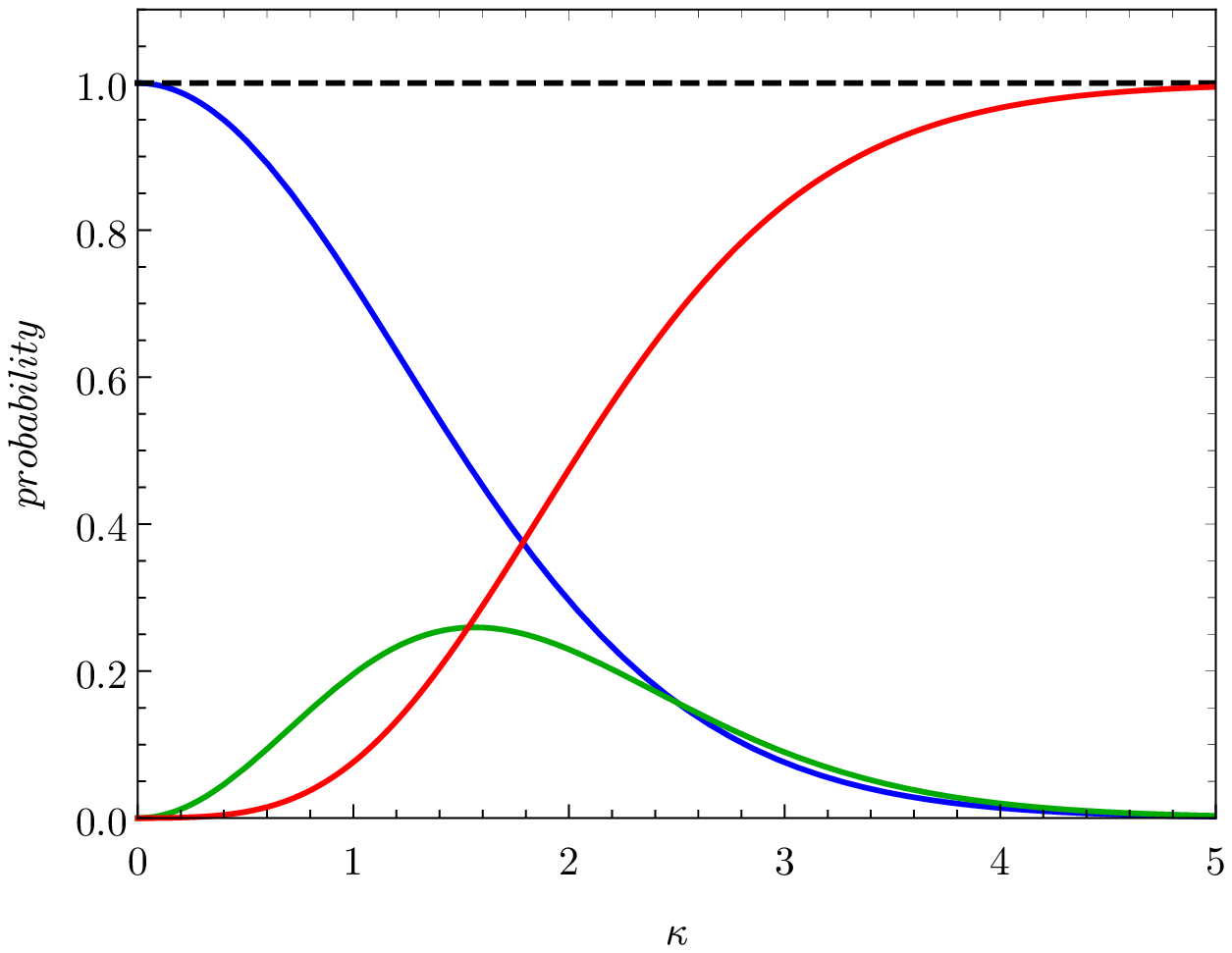} \ \ \ 
		\includegraphics[width=5cm]{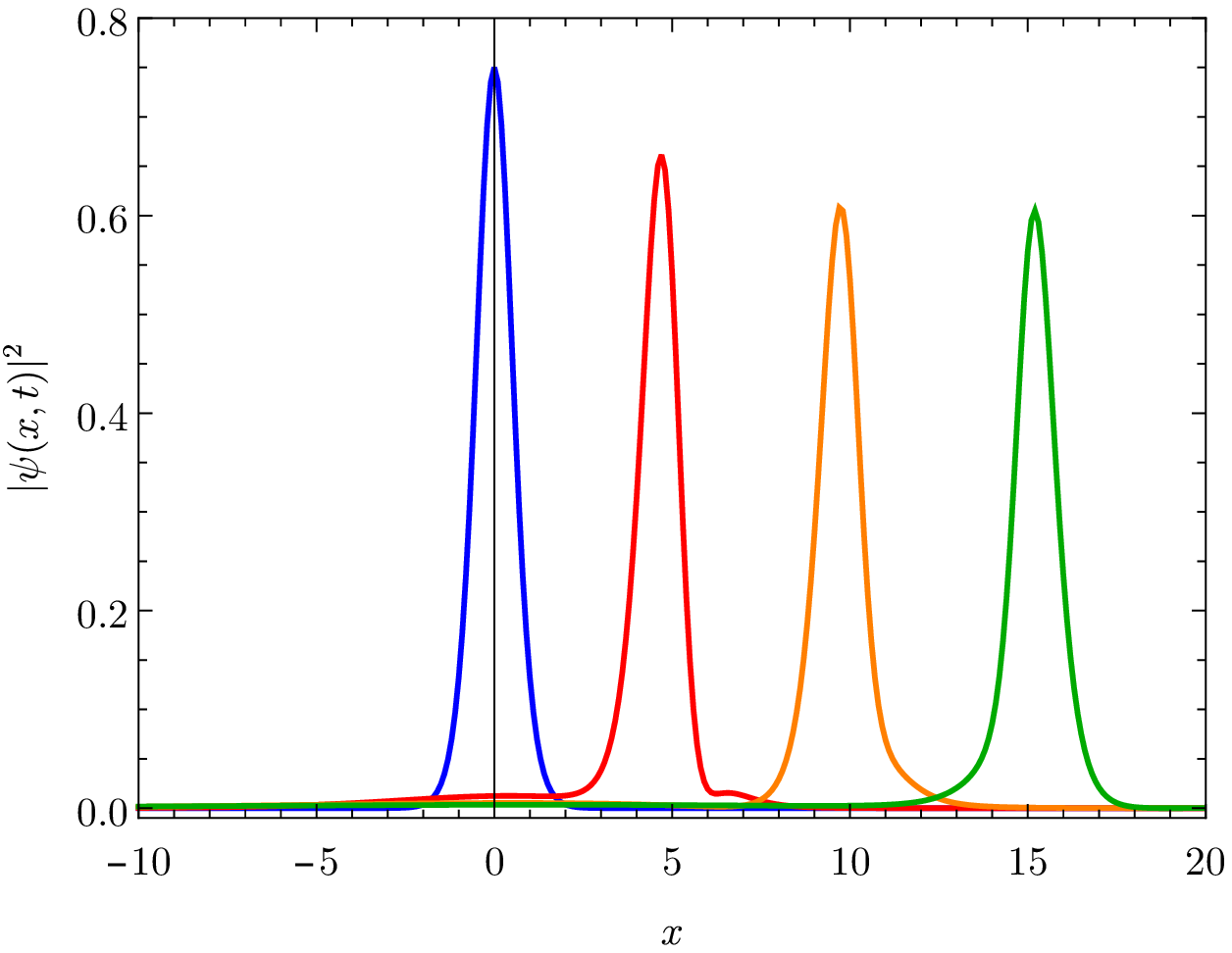} \\
		\caption{Left plot: Probability that a particle initially in the bound state of the Poschl-Teller potential with $\lambda=2$ stays
			trapped as the well starts to move with constant velocity (blue and green curves); the red curve is the probability that the particle ends up in a state of the continuum. Right plot: Probability density at four different times $t=0,5,10,15$ for a particle initially in the bound state of the static potential with $\lambda=2$.  We have used $a=m=\hbar=\kappa=1$. }
		\label{Fig_2}
	\end{center}
\end{figure}

The plots in Figs.~\ref{Fig_2}  are the analogous of the plots in Figs.~\ref{Fig_1}, but for a potential with $\lambda=2$;  
in this case the potential possess two bound states and, as the potential starts to move, the probability of exciting 
the first excited state of the well grows up to a maximum (for $\kappa \approx 1.5$) and then decreases (see the left plot in Fig.~\ref{Fig_2}). 
The conservation of total probability is verified numerically to hold.

The location of the maximum of the probability of exciting a different bound state approximately corresponds to absorbing 
the kinetic energy of the particle and make a transition to the other bound state
\begin{eqnarray}
- \frac{\hbar^2}{2 m a^2} \left( \mu^2 - \lambda^2 \right) = \frac{1}{2} m v^2
\end{eqnarray}
or equivalently
\begin{eqnarray}
\kappa = \sqrt{\lambda^2 -\mu^2}
\end{eqnarray}
For the case in the left plot of Fig.~\ref{Fig_2}, $\lambda = 2$ and $\mu=1$, and therefore the maximum corresponds to 
$\kappa = \sqrt{3}$.

\begin{figure}
	\begin{center}
		\bigskip\bigskip\bigskip
		\includegraphics[width=6cm]{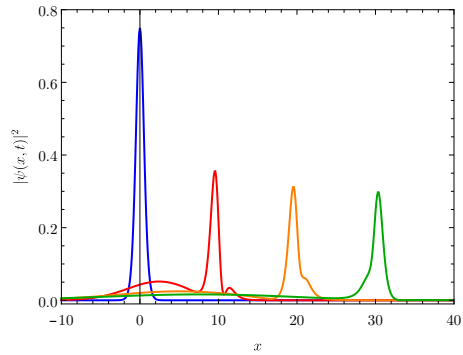} \\
		\caption{Probability density at four different times $t=0,5,10,15$ for a particle initially in the bound state of the static
			potential with $\lambda=2$.  We have used $a=m=\hbar=1$ and $\kappa=2$.  }
		\label{Fig_3}
	\end{center}
\end{figure}

In Fig.~\ref{Fig_3} we display the time evolution of the wave function for $\lambda=2$ but for $\kappa=2$, at which the components of
the bound states are comparable: in this case one can appreciate the asymmetric time-dependent shape of the peak, which reflects the fact that the  particle is not in a stationary state.

\subsection{Simple harmonic oscillator}

The simple harmonic oscillator 
\begin{eqnarray}
V(x) = \frac{1}{2} \ m \omega^2 x^2
\end{eqnarray}
is possibly the most important example of quantum mechanical problem for which exact solutions are known.

In this case the spectrum is discrete, with bound states of energy
\begin{eqnarray}
E_n = \hbar \omega \left(n+\frac{1}{2}\right) \ \ \ , \ \ \ n=0,1,\dots 
\end{eqnarray}
and with eigenfunctions 
\begin{eqnarray}
\psi_n(x) = \frac{1}{ \sqrt{2^n n!}} \left({\frac{m \omega }{\pi \hbar }}\right)^{1/4} e^{-\frac{m x^2 \omega }{2 \hbar }}
H_n\left(x \sqrt{\frac{m \omega }{\hbar}}\right)
\end{eqnarray}
where $H_n(x)$ is the Hermite polynomial of order $n$.

We define the dimensionless parameter 
\begin{eqnarray}
\kappa \equiv \frac{m v^2}{2\hbar \omega}
\end{eqnarray}
representing  the ratio between the kinetic energy associated with the motion of the well and a quanta of energy $\hbar \omega$.

Assuming that the particle is initially in the ground state of the static potential, the amplitudes can be obtained explicitly;
the first few amplitudes are
\begin{eqnarray}
\mathcal{Q}_{0,0}(\kappa) = e^{-\kappa /2} \ \ \ &,&  \ \ \ 
\mathcal{Q}_{0,1}(\kappa) = -i e^{-\kappa /2} \sqrt{\kappa }  \nonumber \\
\mathcal{Q}_{0,2}(\kappa) = -\frac{e^{-\kappa /2} \kappa }{\sqrt{2}}  \ \ \ &,& \ \ \ 
\mathcal{Q}_{0,3}(\kappa) = \frac{i e^{-\kappa /2} \kappa^{3/2}}{\sqrt{6}} \nonumber
\end{eqnarray}

In this case the maximum of the probability $\left| \mathcal{Q}_{0n}\right|^2$ of exciting the state $n$ 
corresponds a velocity given by
\begin{eqnarray}
\frac{m v^2}{2} = \hbar \omega n 
\end{eqnarray}
or equivalently
\begin{eqnarray}
\kappa=n 
\label{sho_cond}
\end{eqnarray}

\begin{figure}
	\begin{center}
		\bigskip\bigskip\bigskip
		\includegraphics[width=6cm]{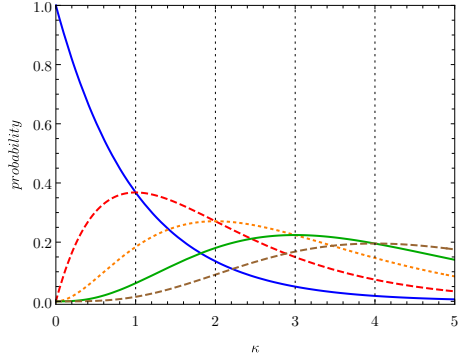} 
		\caption{Probability of transition from the ground state to an excited state for the simple harmonic 
			oscillator, due to a sudden movement. The vertical lines correspond to the condition (\ref{sho_cond}).}
		\label{Fig_SHO}
	\end{center}
\end{figure}

The probabilities of transition from the ground state of the SHO to an excited state, due to a sudden movement of the
well, are plotted in Fig.~\ref{Fig_SHO}. The vertical lines in the plot correspond to the condition (\ref{sho_cond}).
It is worth noticing that at $\kappa = n$, $|\mathcal{Q}_{0,n}(n)|^2 = |\mathcal{Q}_{0,n-1}(n)|^2$, for $n>1$. The proof of this property can be found in  Appendix \ref{appA}.

\subsection{Hydrogen atom}
\label{sub:hyd}

Let us now consider a hydrogen atom, initially at rest in the ground state, which is suddenly kicked and the proton 
starts to move with constant velocity $\vec{v}$.

The wave function of the bound states of the (static) hydrogen atom read
\begin{eqnarray}
\psi_{nlm}(r,\theta,\phi) &=& R_{nl}(r) Y_l^m(\theta ,\phi ) \nonumber \\
&=&  2^{l+1} e^{-\frac{r}{a_0 n}} \sqrt{\frac{(-l+n-1)!}{a_0^3 n^4 (l+n)!}}
\left(\frac{r}{a_0 n}\right)^l L_{-l+n-1}^{2 l+1}\left(\frac{2 r}{n a_0}\right) Y_l^m(\theta ,\phi )
\end{eqnarray}
with $n \geq 1$, $0 \leq l \leq n-1$ and $|m| \leq l$.

Based on our general discussion, the amplitude for the transition from the ground state to any excited state reads~\footnote{Since the
	quantum number for the third component of angular momentum needs to vanish,$m=0$, we express the amplitudes as functions of $n$ and $l$ alone.}
\begin{eqnarray}
\mathcal{Q}_{1,0;n,l} &=& \int e^{-i \frac{\mu v r \cos\theta}{\hbar}} \psi^\star_{n,l,0}(\vec{r}) \psi_{1,0,0}(\vec{r}) d^3r
\end{eqnarray}
where $\mu$ here is the mass of the electron.

Using the partial wave decomposition of a plane wave
\begin{eqnarray}
e^{i \frac{\mu v r  \cos\theta}{\hbar}} = \sum_{l=0}^\infty (2l+1) \ i^l \ j_l\left(\frac{\mu v r}{\hbar}\right) \ P_l(\cos\theta) 
\nonumber 
\end{eqnarray}
one can reduce the expressions to a one dimensional integral
\begin{eqnarray}
\mathcal{Q}_{1,0;n,l}(\kappa) &=&  \sqrt{2l+1} i^l \int_0^\infty j_l\left(\frac{\mu v r}{\hbar}\right) R_{10}(r) R_{nl}(r) r^2 dr
\end{eqnarray}

We have calculated explicitly the amplitudes for $1 \leq n \leq 10$; their expressions 
(not reported here) depend uniquely on the dimensionless parameter
$\kappa \equiv \frac{\hbar v}{e^2/4\pi \epsilon_0} =  \mu v a_0/\hbar$ ($a_0$ is the Bohr radius).
Just to get an idea, $\kappa = 1$ corresponds to a speed $v \approx 0.007892 c$; using the root mean square speed for
hydrogen gas $v_{rms} = \sqrt{3 k_B T/M}$ we can associate a temperature $T \approx 2.2 \times 10^8 \ K$. To obtain a sizeable 
ionization effect on the atoms of a gas trapped in a container by means of the elastic collisions of the individual atoms  with the walls of the container, one should reach incredibly high temperatures~\footnote{Of course, this argument is only qualitative since the boundary conditions for that problem would be different and the wave functions for the moving and static systems would not be related by eq.~(\ref{subst}).}.

Using these expressions we can calculate exacly the probability of a transition $(1,0) \rightarrow (n,l)$, with 
$n \leq N$, due to a sudden movement with velocity $\vec{v}$~\footnote{Notice that for the hydrogen atom there is an
	infinite number of bound states.}
\begin{eqnarray}
P_{n \leq N} = \sum_{n=0}^N \sum_{l=0}^{n-1} \left| \mathcal{Q}_{1,0;n,l}(v) \right|^2
\end{eqnarray}

Of course the probability of not ionizing the atom corresponds to using $N\rightarrow \infty$ in the expression above.

At small velocities ($\kappa \ll 1$) we may obtain the leading behavior of this probability
\begin{eqnarray}
P_{n \leq 6} &\approx& 1 -0.302617 \kappa^2-0.576334 \kappa ^4+\dots \nonumber \\
P_{n \leq 7} &\approx& 1 -0.297702 \kappa^2-0.572154 \kappa ^4 +\dots \nonumber \\
P_{n \leq 8} &\approx& 1 -0.294468 \kappa^2-0.569285 \kappa ^4 +\dots \nonumber \\
P_{n \leq 9} &\approx& 1 -0.292225 \kappa^2-0.567241 \kappa ^4 +\dots \nonumber \\
P_{n \leq 10} &\approx& 1 -0.290603 \kappa^2 - 0.565735 \kappa ^4 +\dots
\end{eqnarray}

The coefficients of the contribution of order $\kappa^2$ form a nice monotonic sequence, so that one can use extrapolation
to estimate its value for $N \rightarrow \infty$ accurately. Moreover, since these coefficients receive contributions only from 
the transition $(1,0) \rightarrow (n,1)$, a larger number of coefficients can be calculated with limited effort.

We have used Richardson extrapolation on the sequence of the first $20$ coefficients obtaining 
\begin{equation}
\begin{split}
 - 0.&28341221595516952094 - 0.78146725925265723860 \ \frac{1}{N^2} \\
&+ 0.78146725925251190251 \ \frac{1}{N^3} +\dots  \nonumber
\end{split}
\end{equation}
showing that the ionization probability for the hydrogen atom goes as
$0.28341221595516952094 \  \kappa^2$ for $\kappa \rightarrow 0$.

This result can be confirmed by using the wave functions of the continuum, $\Psi_{klm}(\vec{r})$.
In this case
\begin{eqnarray}
\mathcal{P}(k) &=& \int e^{-i \frac{\mu v r \cos\theta}{\hbar}} \Psi^\star_{klm}(\vec{r}) \psi_{1,0,0}(\vec{r}) d^3r
\end{eqnarray}
is the amplitude for the transition from the ground state to the continuum (i.e. its modulus square is the probability of ionizing the atom).

The direct calculation of the ionization probability requires using the continuum wave functions. These wave functions
are reported for instance in Ref.~\cite{Peng10} and read:
\begin{eqnarray}
\Psi_{klm}(r,\theta,\phi) &=& Y_{lm}(\theta,\phi) \ \frac{\phi_{kl}(r)}{r} \equiv Y_{lm}(\theta,\phi) \ R_{l}(k,r)
\end{eqnarray}
where (note that the radial wave functions $R_{l}(k,r)$ obey the $k/(2 \pi)$ normalization)
\begin{equation}
\begin{split}
\phi_{kl}(r) &= \frac{2^{l+1}  (k r)^{l+1}}{a_0 r (2 l+1)!} e^{\frac{\pi }{2 a_0 k}-i k r} \left| \Gamma\left(l-\frac{i}{k a_0}+1\right)\right| \\
&\times  ~_1F_1\left(l+\frac{i}{k a_0}+1;2 (l+1);2 i k r\right) \nonumber 
\end{split}
\end{equation}

Using these wave functions we obtain the exact expression for the leading  behavior of the ionization probability as $\kappa \rightarrow 0$
\begin{equation}
\begin{split}
\int_0^\infty & \left| \mathcal{P}(k) \right|^2 \frac{dk}{2\pi} \approx  \\
& \kappa^2 \ \int_0^\infty \frac{256 \pi  u
	\left(\frac{u+i}{-u+i}\right)^{-i/u} \left(-1+\frac{2 i}{u+i}\right)^{i/u}
	\left(\coth \left(\frac{\pi}{u}\right)+1\right)}{3 \left(u^2+1\right)^5} \frac{du}{2\pi} + O(\kappa^4) \nonumber \\
&\approx 0.28341221595516952089 \ \kappa^2 \nonumber 
\end{split}
\end{equation}

The value obtained using the extrapolation of the complementary probabilities provides a very accurate estimate 
(with an error approximately of $4.7 \times 10^{-20}$). 

\section{Conclusions}
\label{concl}
We have extended the one dimensional model of Granot and Marchewka in Ref.~\cite{Granot09} 
for an atom displacing with a moving tip (represented by a Dirac delta function) to a number of
potentials with different spectrum (both discrete and mixed) and in one and three dimensions.
Our calculations are based on a spectral decomposition, rather than on the direct use of the 
propagator (as done in Ref.~\cite{Granot09}) and, for the case discussed in \cite{Granot09} 
we reproduce the probability that the particle stays trapped calculated by Granot and Marchewka.

The remaining examples that we discuss present new and interesting features, not found in the example 
considered in Ref.~\cite{Granot09}: for instance, for the case of P\"oschl-Teller potentials 
we show that the reflected peak moving with velocity $2v$  found in Ref.~\cite{Granot09}  is absent, 
due to the reflectionless nature of PT potentials; moreover, for the case of potentials with more 
than one bound state, there is a probability that the particle gets to an excited bound state, rather 
than to the continuum and we have found a simple criterium based on energy conservation to identify 
the maxima of this probability. Finally, we have calculated {\sl exactly}  the leading contribution
in the velocity $v$ to the probability that a hydrogen atom gets ionized due to a sudden movement 
of the proton.

\appendix

\section{Amplitudes for the simple harmonic oscillator}
\label{appA}

We can understand this result in terms of the creation and annihilation operators 
$\hat{a}$ and $\hat{a}^\dagger$; calling $|n\rangle$ an eigenstate of the Hamiltonian 
of the simple harmonic oscillator we have
\begin{eqnarray}
\hat{a}^\dagger\hat{a}|n\rangle &=& n|n\rangle \nonumber\\
\hat{a}|n\rangle &=& \sqrt{n}|n-1\rangle \nonumber\\
\hat{a}^\dagger|n\rangle &=& \sqrt{n+1}|n+1\rangle. \nonumber 
\end{eqnarray}

Following Ref.~\cite{Fernandez95} we define 
\begin{eqnarray}
I_{mn} \equiv \langle m | U | n \rangle \nonumber
\end{eqnarray}
where $\hat{U} = e^{-\alpha \hat{x}}$ and $\hat{x} = \sqrt{\frac{\hbar}{2m\omega}}\left(\hat{a} + \hat{a}^{\dagger}\right)$.

Then
\begin{equation*}
\begin{split}
I_{0,n} & = \langle 0|\hat{U}|n\rangle = \langle 0|\hat{U}\frac{\hat{a}^\dagger}{\sqrt{n}}|n-1\rangle = \frac{1}{\sqrt{n}}\langle 0|\left(\hat{a}^\dagger-\alpha\sqrt{\frac{\hbar}{2m\omega}}\right)\hat{U}|n-1\rangle  \\
& = -\frac{\alpha}{\sqrt{n}}\sqrt{\frac{\hbar}{2m\omega}}\langle 0|\hat{U}|n-1\rangle = -\frac{\alpha}{\sqrt{n}}\sqrt{\frac{\hbar}{2m\omega}} I_{0,n-1}.
\end{split}
\end{equation*}

On the other hand $I_{0,0} = e^{\alpha^2 \hbar/4 m\omega}$ and for $\kappa/2 = -\frac{\hbar}{4m\omega}\alpha^2$ we obtain 
\begin{equation*}
I_{0,n}(\kappa) = -i\sqrt{\frac{\kappa}{n}}I_{0,n-1}(\kappa)
\end{equation*}
from which the property
\begin{equation*}
|I_{0,n}(n)|^2 = |I_{0,n-1}(n)|^2
\end{equation*}
follows.

\section*{Acknowledgements}
The research of P.A. was supported by Sistema Nacional de Investigadores (M\'exico).

\end{document}